\documentclass[letterpaper]{article}
\usepackage{times}
\usepackage{helvet}
\usepackage{courier}
\usepackage{fixbib}
\usepackage{amsmath}
\usepackage{amssymb}
\usepackage{color}
\usepackage{epsf}
\usepackage{color}
\usepackage{graphicx}
\usepackage{epsfig}
\usepackage{fancybox}
\usepackage{amssymb}
\usepackage{amsmath}
\usepackage{multicol}
\usepackage{hhline}
\usepackage{xspace,epic,eepic,graphicx}
\usepackage{latexsym}
\usepackage[all]{xy}
\usepackage{enumerate}

\mathchardef\myhyphen="2D

\newcommand{\boxtheorem}{\hfill $\blacksquare$\vspace{1mm}}
\newcommand{\ignore}[1]{}
\newcommand{\nit}[1]{{\it #1}}

\DeclareMathAlphabet{\mathpzc}{OT1}{pzc}{m}{it}

\newcommand{\eat}[1]{}
\newcommand{\mc}[1]{\mathcal{ #1}}

{\vskip \abovedisplayskip \refstepcounter{lemmaA-counter}%
\noindent {\bf Lemma A.\arabic{lemmaA-counter}.}}%

{\vskip \abovedisplayskip \refstepcounter{definitionA-counter}%
\noindent {\bf Definition A.\arabic{definitionA-counter}.}}%

\newcommand{\defproof}[2]{{\noindent\bf Proof of #1:\
}#2 \boxtheorem\\ \vspace{2mm}}

\newcommand{\hproof}[1]{{\noindent\bf Proof:\
}#1 \boxtheorem\\ }

\newcommand{\bblue}[1]{\textcolor{blue}{#1}}

\newcommand{\dproof}[1]{{\noindent\bf Proof:\
}#1 \boxtheorem}

\newcounter{theorem-counter}
\newcounter{corollary-counter}
\newcounter{lemma-counter}
\newcounter{definition-counter}
\newcounter{example-counter}
\newcounter{proposition-counter}
\newcounter{remark-counter}
\newcounter{definitionA-counter}
\newcounter{lemmaA-counter}
\newcounter{propositionA-counter}
{\vskip \abovedisplayskip \refstepcounter{theorem-counter}%
\noindent {\bf Theorem \arabic{theorem-counter}.}}%

{\vskip \abovedisplayskip \refstepcounter{corollary-counter}%
\noindent {\bf Corollary \arabic{corollary-counter}.}}%

{\vskip \abovedisplayskip \refstepcounter{lemma-counter}%
\noindent {\bf Lemma \arabic{lemma-counter}.}}%

\newenvironment{definition}%
{\vskip \abovedisplayskip \refstepcounter{definition-counter}%
\noindent {\bf Definition \arabic{definition-counter}.}}%

\newenvironment{example}%
{\vskip \abovedisplayskip \refstepcounter{example-counter}%
\noindent {\bf Example \arabic{example-counter}.}}%

\newenvironment{proposition}%
{\vskip \abovedisplayskip \refstepcounter{proposition-counter}%
\noindent {\bf Proposition \arabic{proposition-counter}.}}%

{\vskip \abovedisplayskip \refstepcounter{remark-counter}%
\noindent {\bf Remark \arabic{remark-counter}.}}%

\title{\bf Causes for Query Answers from Databases, Datalog Abduction and View-Updates: The Presence of Integrity Constraints}
\date{}
\author{{\bf Babak Salimi}\thanks{{\small \tt bsalimi@cs.washington.edu.} Now at University of Washington. This work was done while at Carleton University.} \ \ and \ \ {\bf Leopoldo Bertossi}\thanks{{\small \tt bertossi@scs.carleton.ca}. Contact author.}
\\
Carleton University, \ School of Computer Science, \
Ottawa, Canada.
}

\begin{document}
\maketitle
\pagestyle{plain}
\thispagestyle{empty}

\begin{abstract}
Causality has been recently introduced in databases, to model, characterize and possibly compute causes
for query results (answers). Connections between query-answer causality, consistency-based diagnosis, database repairs
(wrt. integrity constraint violations), abductive diagnosis and the view-update problem have been established. In this work we further investigate connections between
query-answer causality  and abductive diagnosis and the view-update problem. In this context, we also define and investigate the notion of query-answer causality in the presence of integrity constraints.
\end{abstract}

\section{Introduction}

 Causality is a deep subject that appears at the foundations of many scientific disciplines; and also something we want to represent and compute to deal with {\em uncertainty} of data, information and theories.  In data management in particular, there is a need to represent, characterize
 and compute  causes that explain why certain query results are obtained or not, or why natural semantic conditions, such
 as integrity constraints, are not satisfied. Causality can also
 be used to explain the contents of a view, i.e. of a predicate with virtual
 contents that is defined in terms of other physical, materialized relations (tables).

Most of the work on causality by the computer science community has been done in the context of knowledge representation, but
 little has been said about causality in data management.
 This
 work is about causality as defined for queries posed to relational databases.

The notion of causality-based explanation for a query result   was introduced in \cite{Meliou2010a}, on the basis of the deeper concepts of {\em counterfactual} and {\em actual causation}. This approach
 can be traced back to
\cite{Halpern05}. We will refer to this notion
as {\em query-answer causality} (or simply, {\em QA-causality}). Under this approach, explanations for query answers
are provided in terms causes for query answers; and these causes are ranked according to their {\em degree of responsibility}, which quantifies the extent by which a
QA-cause contributes to an answer. In \cite{Meliou2011}, {\em view-conditioned causality} (vc-causality) was proposed  as a restricted form of QA-causality, to determine causes for  unexpected
query results, but  conditioned to the correctness of prior knowledge  that cannot be altered by counterfactual tuple-deletions.

In \cite{icdt15}, connections were established between QA-causality and {\em database repairs} \cite{2011Bertossi}, which allowed  to obtain several complexity results for QA-causality related problems.
 A connection between QA-causality and {\em consistency-based diagnosis} \cite{Reiter87} was established
in  \cite{icdt15}, characterizing  causes and responsibilities  in terms of diagnoses, and leading to
new results for QA-causality. In \cite{uai15} connections between QA-causality and {\em abductive diagnosis} \cite{console91,EiterGL97} were presented.

The definition of QA-causality applies to monotone queries \cite{Meliou2010a}, but all complexity and algorithmic results in \cite{Meliou2010a,icdt15} have
been  for first-order monotone queries, mainly conjunctive queries. However, QA-causality can be applied to Datalog queries \cite{Abiteboul95}, which are also monotone, but may contain recursion.
Oh the other hand,
abductive diagnosis can be done on top of Datalog specifications, leading to Datalog-abduction, for which there are known complexity results \cite{EiterGL97}. Actually, in \cite{uai15} computational
and complexity results were obtained for Datalog QA-causality from a connection with Datalog-abduction.
In this work we further exploit this connection to obtain new complexity results for Datalog QA-causality.

In \cite{uai15}, connections are reported between QA-causality and the classical  {\em view-update problem} in databases, which is
   about updating a database through views \cite{Abiteboul95}. One wants the base relations (also called ``the source database") to change in a minimal way
   while still producing the view updates. When only deletions are performed on monotone views, we have the {\em delete-propagation problem}, from
views to base relations \cite{BunemanKT02,Kimelfeld12a,Kimelfeld12b}. This is the one considered in this work.

In \cite{uai15}, several connections between QA-causality and the delete-propagation problem were established and used to obtain new results for the former.
 In this work we obtain new results for {\em view-conditioned causality} from this connection.

 We define and investigate the notion of query-answer causality in the presence of integrity constraints. The latter are logical dependencies between database tuples that,
 under the assumption that they are satisfied, should have an effect on determining causes for a query answer. We propose a notion of cause that takes them into account.

\ignore{All the results formally stated in this work and not attributed to any author are new. Some already established results are formally stated (and properly attributed),  to make use of them.}
A slightly extended version of this work, with more examples, can be found in \cite{flairsExt}.

\section{Preliminaries}\label{sec:prel}

We  consider relational database schemas, $\mathcal{S} = (U, \mc{P})$, with $U$ a possibly infinite
data domain, and $\mc{P}$  a finite set of {\em database predicates} of fixed arities. We may use implicit built-in predicates, e.g. $\neq$. Schema $\mc{S}$ determines a language, $\mc{L}(\mc{S})$, of first-order (FO) predicate logic. An instance $D$
for $\mathcal{S}$ is a finite set of ground atomic formulas, a.k.a. {\em tuples},  $P(c_1,..., c_n)$, with $c_i \in U$, and $P \in \mc{P}$ is not a built-in.

A {\em conjunctive query}  ({CQ}) is a formula   of $\mc{L}(\mc{S})$ of the form \ $\mc{Q}(\bar{x})\!: \exists \bar{y}(P_1(\bar{s}_1) \wedge \cdots \wedge P_m(\bar{s}_m))$,
with the $P_i(\bar{s}_i)$ atomic formulas, i.e.  $P_i \in \mc{P}$ or is a built-in, and the $\bar{s}_i$ are sequences of terms, i.e. variables or constants of $U$. The $\bar{x}$ in  $\mc{Q}(\bar{x})$ shows
all the free variables in the formula, i.e. those not appearing in $\bar{y}$.  A sequence $\bar{c}$ of constants is an answer to query $\mc{Q}(\bar{x})$ if $D \models \mc{Q}[\bar{c}]$, i.e.
the query becomes true in $D$ when the variables are replaced by the corresponding constants in $\bar{c}$. We denote the set of all answers to a query $\mc{Q}(\bar{x})$ with $\mc{Q}(D)$.
A conjunctive query is {\em Boolean} (a  {BCQ}), if $\bar{x}$ is empty, i.e. the query is a sentence, in which case, it is true or false
in $D$, denoted by $D \models \mc{Q}$ (or $\mc{Q}(D) =\{\nit{true}\}$) and $D \not\models \mc{Q}$ (or $\mc{Q}(D) =\{\nit{false}\}$), respectively.

Query $\mc{Q}$ is {\em monotone} if for any instances $D_1 \subseteq D_2$, \ $\mc{Q}(D_1) \subseteq \mc{Q}(D_2)$. {CQ}s and unions of  {CQ}s ({UCQ}s) are monotone, so as (possibly not FO) Datalog
queries \cite{Abiteboul95}. {\em We consider only monotone queries.}

An {\em integrity constraint} (IC) is a sentence $\varphi \in \mc{L}(\mc{S})$. Then, given an instance $D$ for schema $\mc{S}$, it may be true or false
in $D$ (denoted $D \models \varphi$, resp. $D \not \models \varphi$). Given a set $\Sigma$ of ICs, a database instance $D$ is {\em consistent} if $D \models \Sigma$; otherwise it is said to be {\em inconsistent}.
In this work we assume that sets of ICs are always finite and logically consistent.

A particular class of  ICs  is formed by {\em inclusion dependencies} ( {IND}s), which are sentences of the form $\forall \bar{x} (P_i(\bar{x}) \rightarrow \exists \bar{y} P_j(\bar{x}',\bar{y}))$,
with $\bar{x}' \cap \bar{y} =\emptyset, \bar{x}' \subseteq \bar{x}$. Another special class of ICs is formed by {\em functional dependencies} (FDs). For example,
 $\psi\!: \forall x \forall y \forall z (P(x, y) \land P(x, z) \rightarrow y = z)$ specifies that the second attribute of $P$ functionally depends upon the first. Notice that it can be written as
 the negation of a BCQ: \ $\neg \exists x \exists y \exists z(P(x, y) \land P(x, z) \land y \neq z)$.

 A Datalog query (DQ)  $\mc{Q}(\bar{x})$ is a program $\Pi$, consisting of positive definite rules of the form \ $P(\bar{t}) \leftarrow P_1(\bar{t}_1), \ldots,
P_n(\bar{t}_n)$, with the $P_i(\bar{t}_i)$ atomic formulas, that accesses an underlying extensional database $D$ (the facts). In particular, $\Pi$  defines
an answer-collecting predicate $\nit{Ans}(\bar{x})$ by means of a top rule of the form $\nit{Ans}(\bar{x}) \leftarrow P_1(\bar{s}_1), \ldots, P_m(\bar{s}_m)$, where the $P_i$
on the RHS are defined by other rules in $\Pi$ or are database predicates for $D$. Here, the $\bar{s}_i$ are lists of variables or constants, and $\bar{x} \subseteq \bigcup_i \bar{s}_1$.

When $\Pi \cup D \models \nit{Ans}(\bar{a})$, $\bar{a}$ is an answer to query $\Pi$ on $D$. Here, $\models$ means that the RHS belongs to the minimal model of $\Pi \cup D$.
The Datalog query is Boolean (a BDQ) if the top answer-predicate is propositional, with a  definition of the form $\nit{ans} \leftarrow P_1(\bar{s}_1), \ldots, P_m(\bar{s}_m)$ \ \cite{Abiteboul95}.
 {CQ}s can be expressed as \ignore{Datalog queries} DQs.

\section{QA-Causality and its Decision Problems} \label{sec:qa-causality}

Following
\cite{Meliou2010a}, in the rest of this work, unless otherwise stated, we assume that a relational database instance $D$ is split in two disjoint sets, $D=D^n \cup D^x$, where $D^n$ and $D^x$ are the sets of {\em endogenous} and {\em exogenous} tuples,
 respectively. The former are admissible, interesting potential causes for query answers; but not the latter. {\em In the rest of this work, whenever a database instance is not explicitly partitioned, we assume all tuples are endogenous.}

 A tuple $\tau \in D^n$ is  a
{\em counterfactual cause} for an answer $\bar{a}$ to $\mc{Q}(\bar{x})$ in $D$  if $D\models \mc{Q}(\bar{a})$, but $D\smallsetminus \{\tau\}  \not \models \mc{Q}(\bar{a})$.  A tuple $\tau \in D^n$ is an {\em actual cause} for  $\bar{a}$
if there  exists $\Gamma \subseteq D^n$, called a {\em contingency set}, such that $\tau$ is a counterfactual cause for $\bar{a}$ in $D\smallsetminus \Gamma$.  $\nit{Causes}(D, \mc{Q}(\bar{a}))$
denotes the set of actual causes for $\bar{a}$. If $\mc{Q}$ is Boolean, $\nit{Causes}(D, \mc{Q})$ contains the causes for answer $\nit{true}$. We collect all
minimal  contingency sets associated with $\tau \in D^n$: \
$\nit{Cont}(D, \mc{Q}(\bar{a}), \tau):=\{ \Gamma \subseteq D^n~|~ D\smallsetminus \Gamma
\models Q(\bar{a}), D\smallsetminus (\Gamma \cup \{\tau\}) \not \models \mc{Q}(\bar{a}), \mbox{and for all }
 \Gamma'\subsetneqq \Gamma, \ D \smallsetminus (\Gamma' \cup \{\tau\})
\models \mc{Q}(\bar{a}) \}$.

The {\em causal responsibility} of a tuple $\tau$ for answer $\bar{a}$ is $\rho_{_{\!\mc{Q}(\bar{a})\!}}(\tau):=\frac{1}{1 + |\Gamma|}$, with $\Gamma$ a
smallest contingency set for $\tau$. When $\tau$ is not an actual cause for $\bar{a}$, $\rho_{_{\!\mc{Q}(\bar{a})\!}}(\tau) :=0$.

QA-causality can be applied  to \ignore{Datalog queries} DQs, denoting with $\nit{Causes}(D, \Pi(\bar{a}))$ the set of causes
for answer $\bar{a}$.

\begin{figure}
  \centering
  \includegraphics[width=1.8in]{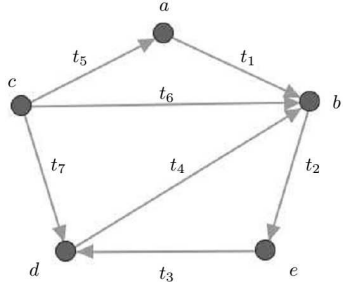}\vspace{-2mm}
  \caption{Graph $G$ for database $D$ in Example \ref{ex:datalogcause}. }\label{fig:gresp}\vspace{-5mm}
\end{figure}

\begin{example} \em \label{ex:datalogcause} \em
Consider the instance $D$ with a single binary relation $E$ as below. $t_1$-$t_7$ are tuple identifiers (ids). Assume all tuples are endogenous.

\vspace{-2mm}
\begin{multicols}{2}
 \begin{center} {\scriptsize \begin{tabular}{l|c|c|} \hline
$E$  & A & B \\\hline
$t_1$ & $a$ & $b$\\
$t_2$& $b$ & $e$\\
$t_3$& $e$ & $d$\\
$t_4$& $d$ & $b$\\
$t_5$& $c$ & $a$\\
$t_6$& $c$ & $b$\\
$t_7$& $c$ & $d$\\ \cline{2-3}
\end{tabular}}
\end{center}

Instance $D$ can be represented as the directed graph $G(\mc{V}, \mc{E})$ in Figure \ref{fig:gresp}, where $\mc{V}$ coincides with the active domain of $D$ (i.e. the set of constants in $E$), and $\mc{E}$ contains an edge\linebreak
  \end{multicols}

\vspace{-6mm}\noindent $(v_1, v_2)$ iff $E(v_1, v_2) \in D$. Tuple ids are used as labels for the  edges in the graph.  For simplicity, we  refer to the  tuples by their
 ids. \ Consider the \ignore{Datalog query} DQ $\Pi$ that collects pairs of vertices of $G$ that are connected through a path, and is formed by the rules: \
$\nit{Ans}(x, y) \leftarrow P(x, y)$. \ \
       $P(x, y) \leftarrow E(x, y)$. and \ \
       $P(x, y) \leftarrow P(x, z), E(z, y)$.

It is easy to see that, $\langle c, e \rangle$ is an answer to query $\Pi$ on $D$. That is,
 $\Pi \cup D \models \nit{Ans}(c, e)$. This is because there are three distinct paths between
 $c$ and $e$ in $G$. All tuples except for $t_3$ are actual causes: $\nit{Causes}(E, \Pi(c, e))=\{t_1, t_2, t_4, t_5, t_6, t_7\}$, because
  all of them contribute to at least one path between $c$ and $e$. Among them, $t_2$ has the
   highest responsibility, because, $t_2$ is a counterfactual cause for the answer, i.e. it has an empty contingency set.
\boxtheorem
\end{example}



The complexity of the computational and decision problems that arise in QA-causality  have been investigated in \cite{Meliou2010a,icdt15}.     For a  Boolean monotone query $\mc{Q}$,
the {\em causality decision problem} (\nit{CDP}) is (deciding about membership of):
$\mc{CDP}(\mc{Q}) := \{(D, \tau)~|~ \tau \in D^n, \mbox{and } \tau \in  \nit{Causes}(D, \mc{Q}) \}$. It is tractable for \nit{UCQ}s \cite{icdt15}.

For a  Boolean monotone query $\mc{Q}$,
the {\em responsibility decision problem} (\nit{RDP}) is (deciding about membership of):
 $\mathcal{RDP}(\mc{Q})=\{(D, \tau, v)~|~ \tau \in D^n, v \in \{0\} \ \cup $
 $\{\frac{1}{k}~|~k \in \mathbb{N}^+\}, $  $D \models \mc{Q}$ \ and \ $\rho_{_{\!\mc{Q}\!}}(\tau) > v  \}$.
It is \nit{ {NP}}-complete for  {UCQ}s \cite{icdt15}.\footnote{All the results are in data complexity.}

\vspace{-1mm}
\subsection{View-conditioned causality}

 Consider  a query $\mc{Q}$ with $\mc{Q}(D)=\{\bar{a}_1, \ldots, \bar{a}_n\}$. Fix an answer, say $\bar{a}_1 \in \mc{Q}(D)$, while the other answers  will be used as a condition on
 $\bar{a}_1$'s causality. Intuitively, $\bar{a}_1$ is somehow unexpected, we look for causes, but considering the other answers as ``correct", which has the effect of
 reducing the spectrum of contingency sets, by keeping $\mc{Q}(D)$'s extension fixed, as a {\em view extension}, modulo $\bar{a}_1$ \ \cite{Meliou2011}.
 More precisely, if $V:= \mc{Q}(D) \smallsetminus \{\bar{a}_1\}$: \
 (a) Tuple $\tau \in D^n$ is a {\em view-conditioned counterfactual cause} ({vcc-}cause) for  $\bar{a}_1$ (in $D$ wrt. $V$) if  $D \smallsetminus \{\tau\} \not \models \mc{Q}(\bar{a}_1)$, but  $\mc{Q}(D \smallsetminus\{\tau\}) =
 V$. (b) Tuple $\tau \in D^n$  is a {\em view-conditioned actual cause} ({vc-}cause) for  $\bar{a}_1$ if there  exists a contingency set, $\Gamma \subseteq D^n$,
 such that $\tau$ is a   {vcc}-cause for $\bar{a}_1$ in $D \smallsetminus \Gamma$ wrt. $V$. \ (c)
 $\nit{vc\mbox{-}\!Causes}(D, \mc{Q}(\bar{a}_1))$ denotes the set of all  {vc-}causes for $\bar{a}_1$. \ (d)
  The {\em {vc-}causal responsibility} of a tuple $\tau$ for answer $\bar{a}_1$ is  $\nit{vc\mbox{-}}\rho_{_{\!\mc{Q}(\bar{a}_1)\!}}(\tau) := \frac{1}{1 + |\Gamma|}$, where $|\Gamma|$ is the
size of the smallest contingency set that makes $\tau$ a vc-cause for  $\bar{a}_1$.

Clearly, $ \nit{vc\mbox{-}\!Causes}(D, Q(\bar{a})) \subseteq \nit{Causes}(D, Q(\bar{a}))$,  but not necessarily the other way around.

\vspace{-.03cm}
\begin{definition}   \label{def:mracpV}
(a) The {\em  {vc-}causality decision problem} ({VCDP}) is about membership of
$\mc{VCDP}(\mc{Q})$ $=\{(D, \bar{a}, \tau)~|~  \bar{a} \in \mc{Q}(D) \ \mbox{and } \tau \in \nit{vc\mbox{-}\!Causes}(D, \mc{Q}(\bar{a}))   \ \}$. \
(b) The {\em  {vc-}causal responsibility decision problem} is about membership of

$\mathcal{VRDP}(\mc{Q})=\{(D,  \bar{a},\tau, v)~|~ \tau \in D^n, v \in \{0\} \ \cup \{\frac{1}{k}~|~k \in \mathbb{N}^+\}, $  $D \models \mc{Q}(\bar{a})$ \ and\\ \hspace*{4.7cm} $\nit{vc\mbox{-}}\rho_{_{\!\mc{Q}\!}}(\tau) > v  \}$. \boxtheorem
\end{definition}

Since leaving the other answers fixed is a strong condition, it makes sense to study the complexity of deciding whether a query answer has a  {vc-}cause or not.

\begin{definition}  \label{def:VCEP}
  For a monotone query $\mc{Q}$, the  {\em  {vc-}cause existence  problem} is (deciding about membership of):
$\mc{VCEP}(\mc{Q})=\{(D, \bar{a})~|~\bar{a} \in \mc{Q}(D) \ \mbox{and }$ $ \nit{vc\mbox{-}\!Causes}(D, \mc{Q}(\bar{a})) \not = \emptyset \ \}.$
\boxtheorem
\end{definition}


\section{Causality and Abduction} \label{sec:abdandcause}

An abductive explanation for an observation is a formula that, together with a background logical theory (a system description), entails the observation.
In database causality we do not have an explicit system description, but just a set of tuples. Something like a system description emerges with a query, and causal relationships between tuples are captured
by the combination of  atoms in it. With a \ignore{Datalog query} DQ \ignore{(in particular, a  {CQ})}, we have a specification in terms of positive definite rules.

A  {\em Datalog abduction problem} \cite{EiterGL97} is of the form $\mathcal{AP}= \langle \Pi, E, \nit{Hyp},$ $ \nit{Obs}\rangle$, where: \ (a) $\Pi$ is a set of Datalog rules,
 (b) $E$ is a  set of ground atoms (the extensional database), (c) the hypothesis, $\nit{Hyp}$,  is a finite set of
ground atoms,  the abducible atoms,\footnote{The hypothesis can be all the possible ground instantiations of {\em abducible predicates}, which
do not appear in rule's LHSs.} and (d) $\nit{Obs}$, the observation, is a finite conjunction of
ground atoms.

The {\em abduction problem} is about computing a subset-minimal $\Delta \subseteq \nit{Hyp}$, such that $\Pi \cup E \cup \Delta \models \nit{Obs}$. In this
case,
$\Delta$ is called an {\em  abductive diagnosis}. So,  no proper subset of $\Delta$ is an abductive diagnosis.
 $\nit{Sol}(\mathcal{AP})$ denotes the set of abductive diagnoses for problem $\mc{AP}$.
Now, a hypothesis $h \in \nit{Hyp}$  is {\em relevant} for $\mathcal{AP}$ if  $h$ contained in at least one diagnosis of $\mathcal{AP}$, otherwise it is {\em irrelevant}. $\nit{Rel}(\mc{AP})$ collects  all relevant hypothesis for $\mc{AP}$. \ A hypothesis $h \in \nit{Hyp}$  is {\em necessary} for $\mathcal{AP}$ if  $h$ contained in all diagnosis of $\mathcal{AP}$. $\nit{Ness}(\mc{AP})$ collects  all the necessary hypothesis for $\mc{AP}$.

The {\em relevance decision problem} (RLDP) is about deciding the membership of:
$\mc{RLDP}(\Pi)=\{(\nit{E, Hyp, Obs}, h)~|~ h \in \nit{Rel}(\mc{AP}), \mbox{with }
\mc{AP}= \langle \Pi, E, \nit{Hyp}, \nit{Obs}\rangle \ignore{\mbox{ and } h \in \nit{Hyp} }\}$.
The {\em necessity decision problem} (NDP) is about deciding the membership of:

$\mc{NDP}(\Pi)=\{(\nit{E, Hyp, Obs}, h)~|~ h \in \nit{Ness}(\mc{AP}), \mbox{with }
\mc{AP}= \langle \Pi, E, \nit{Hyp}, \nit{Obs}\rangle \ignore{\mbox{and } h \in \nit{Hyp}} \}$.

 The following results can be obtained  adapting results in \cite[the. 26]{EiterGL97} and \cite{Friedrich90}:
  For every Datalog program $\Pi$,
$\mc{NDP}(\Pi)$ is in  {\nit{PTIME}} (in data); and, for Datalog programs, $\Pi$, $\mc{RLDP}(\Pi)$
 is {\em  {NP}}-complete.

For
 a \ignore{boolean Datalog query} BDQ  $\Pi$ with  $\Pi \cup D \models \nit{ans}$, the causality decision problem takes the form: \
$\mc{CDP}(\Pi) := \{(D, \tau)~|~ \tau \in D^n, \mbox{and } \tau \in  \nit{Causes}(D, \Pi) \}$.
It turns out that, for  Datalog system specifications,  actual causes for $\nit{ans}$ can be obtained from abductive diagnoses of the associated {\em causal Datalog abduction problem} ({CDAP}): \  $\mathcal{AP}^c:=\langle \Pi, D^x, D^n, \nit{ans}\rangle$, where
 $D^x$ takes the role of the extensional database for $\Pi$. Accordingly, $\Pi \cup D^x$ becomes  the {\em background theory}, $D^n$ becomes the set of {\em hypothesis}, and atom $\nit{ans}$ is the observation.
\vspace{-.1cm}
\begin{proposition}  \label{pro:abdf&cfcaus}
For an instance $D=D^x \cup D^n$ and a \ignore{boolean Datalog query} BDQ $\Pi$, with $\Pi \cup D \models \nit{ans}$, and   its associated CDAP $\mc{AP}^c$, the following hold:
\ (a)
   $ \tau \in D^n$ is an counterfactual cause for $\nit{ans}$  iff $\tau \in \nit{Ness}(\mathcal{AP}^c)$. \ (b)
 $\tau \in D^n$ is an actual cause for $\nit{ans}$  iff $\tau \in \nit{Rel}(\mathcal{AP}^c)$. \boxtheorem
\end{proposition}

\ignore{
\dproof{Part (a) is straightforward. To proof part (b), first assume $\tau$ is an actual cause for
\nit{ans}. According to the definition of an actual cause, there
 exists a contingency set  $ \Gamma \subseteq D^n$ such that $ \Pi \cup D \smallsetminus
  \Gamma \models  \nit{ans}$ but $\Pi \cup D \smallsetminus (\Gamma \cup \{\tau\}) \not \models  \nit{ans}$. This implies
  that there exists a set $\Delta \subseteq D^n$ with $\tau \in \Delta $ such that   $ \Pi \cup \Delta \models \nit{ans}$.
   It is easy to see that  $ \Delta$ is an abductive diagnosis for $\mathcal{AP}^c$. Therefore, $\tau \in \nit{Rel}(\mathcal{AP}^c)$.

Second, assume $\tau \in \nit{Rel}(\mathcal{AP}^c)$. Then there exists a set
 $\mc{S}_k \in \nit{Sol}(\mathcal{AP}^c)=\{s_1 \ldots s_n\}$ such that
 $\mc{S}_k \models \nit{ans}$ with $\tau \in \mc{S}_k$. Obviously, $\nit{Sol}(\mathcal{AP}^c)$ is a
 collection of subsets of $D^n$. Pick a  set $\Gamma \subseteq D^n$ such that for all
  $\mc{S}_i \in \nit{Sol}(\mathcal{AP}^c)$ $i \not = k$, $\Gamma \cap \mc{S}_i \not = \emptyset$ and
    $\Gamma \cap \mc{S}_k =\emptyset$. It is clear that $ \Pi \cup D \smallsetminus (\Gamma \cup \{t\}) \not \models \nit{ans}$ but
     $\Pi \cup D \smallsetminus \Gamma \models \nit{ans}$. Therefore, $\tau$ is an actual cause for \nit{ans}.
To complete the proof we need to show that such $\Gamma$ always exists. This
 can be done by applying
the digitalization technique to construct such $\Gamma$. Since all elements of $\nit{Sol}(\mathcal{AP}^c)$
are subset-minimal, then, for each $\mc{S}_i \in \nit{Sol}(\mathcal{AP}^c)$ with $i \not = k$, there exists a $\tau' \in \mc{S}_i$
 such that $\tau' \not \in \mc{S}_k$.  So, $\Gamma$ can be obtained from the union of differences between each $\mc{S}_i$ ($i \not = k$) and $\mc{S}_k$.
}
}

\begin{example} \label{ex:abdex3}
Consider the instance $D$ with relations $R$ and $S$ as below, and the query $\Pi\!: \ \nit{ans} \leftarrow R(x, y), S(y)$,
which is true in $D$.
Assume all tuples are endogenous.


\begin{center}{\small \begin{tabular}{l|c|c|} \hline
$R$  & A & B \\\hline
 & $a_1$ & $a_4$\\
& $a_2$ & $a_1$\\
& $a_3$ & $a_3$\\
 \hhline{~--}
\end{tabular} \hspace*{1cm}\begin{tabular}{l|c|c|}\hline
$S$  & A  \\\hline
 & $a_1$ \\
& $a_2$ \\
& $a_3$ \\ \hhline{~-}
\end{tabular}}
\end{center}

Here, $\mathcal{AP}^c= \langle \Pi,$ $\emptyset,D,$ $\nit{ans}\rangle$, which
has two (minimal) abductive diagnoses:  $\Delta_1=$ $\{S(a_1),R(a_2, a_1)\}$
  and $\Delta_2=\{S(a_3), R(a_3, a_3)\}$. Then,
$\nit{Rel}(\mc{AP}^c)$ $=$ $\{S(a_3), $ $  R(a_3, a_3), $ $S(a_1), R(a_2, a_1)\}$.
It is clear that the relevant hypothesis are actual causes for $\nit{ans}$.
\boxtheorem
\end{example}

We can use the results mentioned above to obtain new complexity results for Datalog  QA-causality. First, for  the problem of deciding if a tuple is a {\em counterfactual cause} for a query answer.
This is a  tuple that, when removed from the database, undermines the query-answer, without having to remove other tuples, as is the case for {\em actual causes}.
Actually, for each of the latter there may be an exponential number of contingency sets \cite{icdt15}. A counterfactual cause is an actual cause with responsibility $1$.
The complexity of this problem can be obtained from the connection between counterfactual causation and the necessity of hypothesis in Datalog abduction.
\vspace{-1mm}
\begin{proposition}\label{pro:CFDP}
For \ignore{boolean Datalog queries} BDQs $\Pi$, $\mc{CFDP}(\Pi) :=  \{(D, \tau)~|~ \tau \in D^n \mbox{ and } \rho_{_{\!\mc{Q}\!}}(\tau)=1 \}$. is in  {\nit{PTIME}} \ (in data).
 \boxtheorem
\end{proposition}

For \ignore{boolean Datalog queries} BDQs $\Pi$, deciding actual causality, i.e. the problem $\mc{CDP}(\Pi)$, is {\em  {NP}}-complete (in data) \cite{uai15}.
The same problem is tractable for {UCQ}s \cite{icdt15}.
Finally, we establish the complexity of the responsibility problem for \ignore{Datalog queries} DQs.
\begin{proposition}\label{pro:rp}
For \ignore{boolean Datalog queries} BDQs $\Pi$, $\mc{RDP}(\Pi)$ is {\em  {NP}}-complete.
\boxtheorem
\end{proposition}

\section{Causality and View-Updates} \label{sec:delp&cause}

There is a close relationship between QA-causality   and the view-update problem in the form of delete-propagation \cite{Abiteboul95}.

 Let $D$ be a database instance, and $\mc{Q}$ a monotone query. For $\bar{a} \in \mc{Q}(D)$, the {\em minimal-source-side-effect deletion-problem} is about computing a subset-minimal $\Lambda \subseteq D$, such that $\bar{a} \ \notin \mc{Q}(D \smallsetminus \Lambda)$.

 Now, following \cite{BunemanKT02}, let $D$ be a database instance $D$, and $\mc{Q}$ a  monotone query:
(a) For $\bar{a} \in \mc{Q}(D)$, the {\em view-side-effect-free deletion-problem} is about computing a  \ignore{minimum-cardinality} $\Lambda \subseteq D$, such that $\mc{Q}(D) \smallsetminus \{\bar{a}\} = \mc{Q}(D
\smallsetminus \Lambda)$. \ (b) The {\em
view-side-effect-free decision problem} is (deciding
about the membership of):
$\mc{VSEFP}(\mc{Q})= \{(D, \bar{a}) ~|~  \bar{a} \in
\mc{Q}(D), \ \mbox{and exists } D' \subseteq D \mbox{ with }$ $ \mc{Q}(D) \smallsetminus
\{\bar{a}\} = \mc{Q}(D')\} $.
The latter  decision problem  is \nit{NP}-complete for conjunctive queries \cite[theorem 2.1]{BunemanKT02}.


 Consider a relational instance $D$, a view $ \mc{V}$ defined by a monotone query $\mc{Q}$. Then, the virtual view extension, $\mc{V}(D)$, is $\mc{Q}(D)$.  For a tuple $\bar{a} \in {\mc{Q}}(D)$, the delete-propagation problem, in its most general form, is about deleting a set of tuples from $D$, and so obtaining a subinstance $D'$ of $D$, such that
$\bar{a} \notin \mc{Q}(D')$. It is natural to expect that the deletion of $\bar{a}$ from ${\mc{Q}}(D)$
can be achieved through deletions from $D$ of actual causes for $\bar{a}$ (to be in the view extension).
However, to obtain solutions to the different variants of this problem, different combinations  of actual causes must be considered \cite{uai15}.


In particular, in \cite{uai15}, it has been shown that actual causes of $\bar{a}$  with their minimal contingency sets are
in correspondence with the solutions
to the minimal-source-side-effect deletion-problem of $\bar{a}$.

Now,  in order to check if there exists a solution to the view-side-effect-free deletion-problem for  $\bar{a} \in \mc{V}(D)$, it is good enough to check if $\bar{a}$ has a view-conditioned cause. Actually, it holds \cite{uai15}:
\ For an instance $D$, a view $ \mc{V}$ defined by a monotone query $\mc{Q}$ with $\mc{Q}(D)=\{\bar{a}_1, \ldots, \bar{a}_n\}$, and  $\bar{a_k} \in \mc{Q}(D)$, \ignore{\ There is a solution to the view-side-effect-free deletion-problem associated to $\bar{a}$, i.e.} \ $(D, \bar{a}_k) \in \mc{VSEFP}(\mc{Q})$ \ iff \
 $\nit{vc\mbox{-}\!Causes}(D, \mc{Q}(\bar{a}_k)) \not = \emptyset$.

We now consider the complexity of the view-conditioned causality problem  (cf. Definition \ref{def:mracpV}). By appealing to the connection between vc-causality and delete-propaga-tion, we obtain for the {\em {vc-}cause existence problem} (cf. Definition \ref{def:VCEP}):
For  {CQ}s $\mc{Q}$, $\mc{VCEP}(\mc{Q})$ is \nit{{NP}}-complete (in data) \cite{uai15}. A polynomial-time Turing (or Cook) reduction from this problem allows us to obtain the next result
about deciding {vc-}causality (cf. Definition \ref{def:mracpV}).
\begin{proposition} \label{pro:VCcausecausality}
For  {CQ}s $\mc{Q}$, $\mc{VCDP}(\mc{Q})$ is  \nit{{NP}}-complete.
\boxtheorem
\end{proposition}

\ignore{In this result, $\nit{NP}$-hardness is defined in terms of ``Cook (or Turing) reductions" as opposed to many-one (or Karp) reductions.
$\nit{NP}$-hardness
under many-one reductions implies $\nit{NP}$-hardness under Cook
reductions, but the converse, although  conjectured
not to hold, is an open problem. However, for Cook
reductions, it is still true that there is no efficient algorithm
for an $\nit{NP}$-hard problem, unless $P = \nit{NP}$. }

By a (Karp) reduction from this problem, we settle the complexity of the  {vc-}causality responsibility problem for conjunctive queries.

\begin{proposition} \label{pro:VCcauseresponsibility}
For  {CQ}s $\mc{Q}$, $\mc{VRDP}(\mc{Q})$ is  \nit{{NP}}-complete. \boxtheorem
\end{proposition}
These results on vc-causality also hold for  {UCQ}s.

\section{QA-Causality under Integrity Constraints} \label{sec:c&ic}

To motivate  a definition
of QA-causality in the presence of integrity constraints (ICs), we start with some remarks.

{\em Interventions} are at the base of Halpern \& Pearl's  approach to causality \cite{Halpern05}, i.e.
actions on the model that define counterfactual scenarios. In  databases, they take the
form of tuple deletions. If a database $D$ satisfies a prescribed set of integrity constraints (ICs), the
instances obtained from $D$ by tuple deletions, as used to determine causes, should  be expected to satisfy the
ICs.

On a different side, QA-causality  in \cite{Meliou2010a}  is {\em insensitive} to equivalent query
rewriting (as first pointed out in \cite{Glavic11}): QA-causes  coincide
for logically equivalent queries. \ignore{ This property is important since databases engines are accustomed to choose and evaluate the simplest equivalent query to the
one at hand. The fact that QA-causality posses this property guarantees that QA-causes given by this definition are the intended one.}
However, QA-causality might be sensitive to equivalent query rewritings in the presence of ICs, as the following example shows.

\begin{example}\label{ex:ICex1}
Let  $\mc{S}= \{\nit{Dep(DName},$ $\nit{TStaff)},$
$\nit{Course(CName},\nit{LName},\nit{DName)}\}$ be relational schema with inclusion dependency
$$I\!: \ \forall x \forall y \ (\nit{Dep}(x, y) \rightarrow \exists u  \  \nit{Course}(u, y, x));$$ 
and instance $D$ for $\mc{S}$:

\begin{center}
{\small
\hspace*{-0.5cm}
\begin{tabular}{c|c|c|} \hline
\nit{Dep} & \nit{DName} &\nit{TStaff}  \\\hline
$t_1$& \nit{Computing} & \nit{John}   \\
$t_2$& \nit{Philosophy} &  \nit{Patrick}   \\
$t_3$&\nit{Math}  &  \nit{Kevin}   \\
 \hhline{~--} \end{tabular}~~~~~~~~~
 \begin{tabular}{c|c|c|c|} \hline
\nit{Course}  & \nit{CName} & \nit{LName} & \nit{DName} \\\hline
$t_4$&\nit{Com08} & \nit{John}  & \nit{Computing} \\
$t_5$&\nit{Math01} & \nit{Kevin}  & \nit{Math} \\
$t_6$&\nit{Hist02}&  \nit{Patrick}   &\nit{Philosophy} \\
$t_7$&\nit{Math08}&  \nit{Eli}   &\nit{Math}  \\
$t_8$&\nit{Com01}&  \nit{John} &\nit{Computing} \\
 \hhline{~---}
\end{tabular} }
\end{center}

 Clearly, $D \models I$. Now, consider the  CQ that collects the teaching staff who are
lecturing  in the department they are associated with: 
 \begin{eqnarray}
\mc{Q}(\nit{TStaff}) &\leftarrow& \nit{Dep(DName,TStaff}),  \label{eq:heads}\\
&& \nit{Course(CName,TStaff, DName}). \nonumber
\end{eqnarray}

\vspace{-1.5mm}
\noindent Here, \ $\mc{Q}(D) = \{\nit{John}, \nit{Patrick}, \nit{Kevin}\}$. Answer $\langle John \rangle$ has the  actual causes: $t_1$, $t_4$ and $t_8$.
$t_1$ is a counterfactual cause, $t_4$ has a single minimal contingency set $\Gamma_1=\{t_8\}$; and
 $t_8$ has a single minimal contingency set $\Gamma_2=\{t_4\}$.

Now, in the presence of IC $I$, $\mc{Q}$ is equivalent with the following query $\mc{Q}'$: (denoted $\mc{Q} \equiv_{\{I\}}\mc{Q}'$, and meaning they give the same answers for every instance that satisfies $I$)
\vspace{-1mm}
  \begin{eqnarray*}
\mc{Q}'(\nit{TStaff}) &\leftarrow& \nit{Dep(DName,TStaff})).  \label{eq:heads2}
\end{eqnarray*}

\vspace{-1mm}
\noindent In particular, $\langle John \rangle$ is still an answer to $\mc{Q'}$ from $D$. However, on the basis of query $\mc{Q}'$ and instance $D$ alone, there is
single cause, $t_1$, which is also a counterfactual cause.
\boxtheorem
\end{example}


\vspace{-1mm}
\begin{definition} \label{def:causeIC}
Given an instance $D=D^n \cup D^x$ that satisfies a set $\Sigma$ of ICs, i.e. $D\models \Sigma$, and a
monotone query $\mc{Q}$ with $D \models \mc{Q}(\bar{a})$, a tuple $\tau \in D^n$ is an  {\em actual cause for  $\bar{a}$
under $\Sigma$} \ if there is $\Gamma \subseteq D^n$, such that:
\begin{itemize}
\item[(a)]
$ D \smallsetminus \Gamma  \models  \mc{Q}(\bar{a})$, \ and \ (b) \
$ D \smallsetminus \Gamma \models \Sigma$.
\item[(c)] $ D \smallsetminus (\Gamma \cup \{t\}) \not \models  \mc{Q}(\bar{a})$, \ and \ (d) \ $ D \smallsetminus (\Gamma \cup \{t\}) \models \Sigma$.

\end{itemize}
$\nit{Causes}(D, \mc{Q}(\bar{a}), \Sigma)$
denotes the set of actual causes for $\bar{a}$ under $\Sigma$. \boxtheorem
\end{definition}
\begin{example}\label{ex:ICex2} (ex. \ref{ex:ICex1} cont.)  Consider  answer  $\langle John \rangle$
to $\mc{Q}$, for which   $t_4$ was a cause with minimal contingency set $\Gamma_1=\{t_8\}$.
It holds $D \smallsetminus \Gamma_1 \models I$, but
$D \smallsetminus (\Gamma_1 \cup \{ t_2\}) \not \models I$. So, the new definition
does not allow $t_4$ to be an actual cause for answer $\langle John \rangle$ to $\mc{Q}$.
Actually,
$\mc{Q}$ and $\mc{Q'}$ have the same actual causes for answer $\langle John \rangle$ under $I$, namely $t_1$.
\ignore{+++
Now consider a query about all lecturers: \vspace{-1mm}
\begin{eqnarray*}
\mc{Q}''(\nit{LName}) &\leftarrow & \nit{Course(CName, LName, DName}).  \label{eq:heads3}
\end{eqnarray*}

\vspace{-1mm}\noindent Again, $\langle John \rangle$ is an
answer to $\mc{Q}''$, with two actual causes under $I$: $t_4$ and $t_8$. The former with contingency set
$\Gamma_3=\{t_8, t_1\}$.

This result is interesting, because \bblue{this answer have the same actual causes wrt. the original definition of QA-causality. However,
the contingency associated to $t_4$ is $\{t_8\}$. This shows that ICs may affect responsibility even for the answers for which we obtain similar causes wrt. both definitions.}}
\boxtheorem
\end{example}
Since functional dependencies (FDs) are never violated by tuple deletions, they have no effect on the set of causes
for a query answer. Actually, this applies to all {\em denial constraints} (DCs), i.e. of the form
$\neg \forall \bar{x}(A_1(\bar{x}_1) \wedge \cdots \wedge A_n(\bar{x}_n))$, with $A_i$ a database predicate or a built-in.

\begin{proposition}
Given an instance $D$, a monotone query $\mc{Q}$, and a set of ICs $\Sigma$, the following hold:
\begin{itemize}
\item[(a)]  $\nit{Causes}(D, \mc{Q}(\bar{a}), \Sigma) \subseteq
 \nit{Causes}(D, \mc{Q}(\bar{a}))$.

 \item[(b)] $\nit{Causes}(D, \mc{Q}(\bar{a}), \emptyset) = \nit{Causes}(D, \mc{Q}(\bar{a})).$

 \item[(c)] When $\Sigma$ consists of DCs,
 $\nit{Causes}(D, \mc{Q}(\bar{a}), \Sigma) = \nit{Causes}(D, \mc{Q}(\bar{a}))$.

  \item[(d)] For a monotone query $\mc{Q}'$ with
  $\mc{Q}' \equiv_\Sigma \mc{Q}$: \
  $\nit{Causes}(D, \mc{Q}(\bar{a}), \Sigma) = \nit{Causes}(D,$ $\mc{Q}'(\bar{a}),\Sigma).$

  \item[(e)] For a monotone query $\mc{Q}'$ which is minimally contained in $\mc{Q}$ with
  $\mc{Q}' \equiv_\Sigma \mc{Q}$:\footnote{This means $\mc{Q}' \subseteq \mc{Q}$ and there is no  $\mc{Q}''$ with
  $\mc{Q}'' \subsetneqq \mc{Q}'$  and $\mc{Q}'' \equiv_\Sigma \mc{Q}$.}
  $\nit{Causes}(D, \mc{Q}(\bar{a}), \Sigma) = \nit{Causes}(D, \mc{Q}'(\bar{a})).$ \boxtheorem
\end{itemize}
\end{proposition}
\vspace{-.2cm}
Notice that item (e) here relates to the rewriting of the query in Example \ref{ex:ICex1}. Notice that this rewriting  resembles the resolution-based rewritings used
in {\em semantic query optimization} \cite{minker}.

Since {FD}s have no effect on causes, the causality decision problems in the presence of  {FD}s have the
same complexity upper bound as causality without FDs. For example, for $\Sigma$ a set of FDs, $\mc{RDP}(\mc{Q},\Sigma)$, the responsibility problem now under
FDs, is \nit{NP}-complete (as it was without ICs \cite{icdt15}).
However,  when an instance satisfies
a set of  {FD}s, the decision problems may become tractable depending on the query structure. For example,
for the class of {\em key-preserving}  CQs, deciding responsibility over instances that satisfy the key constraints (KCs) is in
{\nit{PTIME}} \cite{Cibele15}.  A KC is a particular kind of FD where some of the predicate attributes functionally determine {\em all}
the others.
Given a set $\kappa$ of KCs, a CQ is key-preserving if, whenever an instance $D$ satisfies $\kappa$, all key attributes
of
base relations involved in $\mc{Q}$ are included among the
attributes of $\mc{Q}$.

By appealing to the connection between vc-causality and delete-propagation \cite{uai15},  vc-responsibility under KCs is tractable (being intractable in general, because the
problem without KCs already is, as shown in Proposition \ref{pro:VCcauseresponsibility}):

\vspace{-1mm}
\begin{proposition} \label{pro:ICcause}
Given a set $\kappa$ of KCs, and a key-preserving {{CQ}} query  $\mc{Q}$, deciding $\mc{VRDP}(\mc{Q},\kappa)$ is in  {\nit{PTIME}}. \boxtheorem
\end{proposition}

New subclasses of (view-defining) CQs for which different variants of delete-propa-gation are tractable are introduced
in \cite{Kimelfeld12a,Kimelfeld12b} (generalizing  those in \cite{Cong95}). The established connections between
delete-propagation and causality should allow us to adopt them for the latter.

QA-causality under ICs can capture  vc-causality:

\ignore{
\vspace{-1mm}
\begin{proposition} \label{pro:vc-ICs}
For a monotone query $\mc{Q}$ with $\mc{Q}(D)=\{\bar{a}_1, \ldots, \bar{a}_n\}$, and  $\bar{a_k} \in \mc{Q}(D)$, there is a set of inclusion dependencies $\Sigma$, such that
$\nit{vc\mbox{-}\!Causes}(D, \mc{Q}(\bar{a}_k))=\nit{Causes}(D, \mc{Q}(\bar{a}), \Sigma)$. \boxtheorem
\end{proposition}  }

\begin{proposition} \label{pro:vc-ICs}
For a conjunctive query $\mc{Q}(\bar{x}) \in L(\mc{S})$, and an instance  $D$ for $\mc{S}$,  with $\mc{Q}(D)=\{\bar{a}_1, \ldots, \bar{a}_n\}$ and  a fixed $k \in \{1, \ldots, n\}$, there is a set of inclusion dependencies $\Sigma$ over schema $\mc{S} \cup \{V\}$, with $V$ a fresh $|\bar{x}|$-ary predicate, and an instance $D'$ for $\mc{S} \cup \{V\}$, such that
$\nit{vc\mbox{-}\!Causes}(D, \mc{Q}(\bar{a}_k))=\nit{Causes}(D', \mc{Q}(\bar{a}), \Sigma)$. \boxtheorem
\end{proposition}

\ignore{
\hproof{Given $D$, consider the instance $D' := D \cup (\mc{Q}(D)\smallsetminus\{\bar{a}_k\})$, where the second disjunct is the extension for predicate $V$. Now, consider the set
of ICs for schema $\mc{S} \cup \{V\}$: \ $\Sigma := \{\forall \bar{x}(V(\bar{x}) \rightarrow \mc{Q}(\bar{x}))\}$. }  }

Deciding causality in the absence of ICs is tractable, but their presence has an
impact on this problem. The following is obtained from Propositions \ref{pro:VCcausecausality} and  \ref{pro:vc-ICs}.

\ignore{
On the other side, deciding causality in the absence of ICs is tractable. However, their presence may have an
impact on this problem.}

\begin{proposition} \label{pro:keyp}
For  CQs $\mc{Q}$ and a set  $\Sigma$ of inclusion dependencies,
$\mc{Q}$, $\mc{CDP}(\mc{Q},\Sigma)$ is {\em  {NP}}-complete. \boxtheorem
\end{proposition}

\ignore{
\hproof{Membership is clear. Now, hardness is established by reduction from the \nit{NP}-complete vc-causality decision problem (cf. Proposition \ref{pro:VCcausecausality}) for a  CQ $\mc{Q}(\bar{x})$ over schema $\mc{S}$.
Now, consider the schema $\mc{S}':= \mc{S} \cup \{V\}$ and the set of ICs $\Sigma$ as in Proposition \ref{pro:vc-ICs}. In order to decide about $(D,\mc{Q}(\bar{a}), \tau)$'s membership of $\mc{VCDP}(\mc{Q})$,
consider the instance $D'$ for $\mc{S}'$ as in Proposition \ref{pro:vc-ICs}, with $\bar{a}$ as $\bar{a}_k$. It holds: \ $(D,\mc{Q}(\bar{a}), \tau) \in \mc{VCDP}(\mc{Q})$ iff
$(D',\mc{Q}(\bar{a}), \tau) \in \mc{CDP}(\mc{Q},\Sigma)$.}
}

Some ICs may be implicative, which makes it tempting to give them a causal semantics. For example,
in \cite{Roy14} and more in the context of interventions for explanations, a ground instantiation, $P_i(\bar{t}_i) \rightarrow P_j(\bar{t}_j)$, of an inclusion dependency is regarded a causal dependency of
$P_j(\bar{t}_j)$ upon $P_i(\bar{t}_i)$. On this basis, a {\em valid intervention}  removes  $P_j(\bar{t}_j)$ whenever  $P_i(\bar{t}_i)$ is removed from the instance.

Giving to ICs a causal connotation is controversial. Actually, according to \cite{Halpern10} logical dependencies
are not causal dependencies {\em per se}. Our approach is consistent with this view. Even more, we should point out that there are different
ways of seeing ICs, and they could have an impact on the notion of cause. For example, according to \cite{Reiter92},  ICs are ``epistemic in nature",
in the sense that rather than being statements about the domain represented by a database (or knowledge base), they are
statement about the {\em contents} of the database, or about what it {\em knows} (cf. \cite{Reiter92} for a discussion).

Abduction has been applied to view-updates \cite{Kakas90}, with ICs on the base relations
\cite{Console95}. On the other side, we have connected QA-causality with both abduction and view-updates. We briefly
illustrate using our ongoing example how the approach in  \cite{Console95} can be used to determine view-updates in the presence of ICs, which should
have an impact on the characterization and computation of causes, now under ICs.
\vspace{-.1cm}
\begin{example} (ex. \ref{ex:ICex1} cont.) Formulated as a view-update problem on a Datalog setting, we have
the query (\ref{eq:heads}) defining an intensional predicate, $\mc{Q}(\nit{TStaff})$. The tuples in the underlying database
are all considered to be abducible. The view-update request is the deletion of $\mc{Q}(\nit{John})$.

According to \cite{Console95}, the potential abductive explanations are maximal subsets $E$ of the original instance $D$, such that
$R$ plus rule (\ref{eq:heads}) does not entail $\mc{Q}(\nit{John})$ anymore. They are: \ $E_1 = D \smallsetminus \{t_1\}$, and $E_2 = D \smallsetminus \{t_4,t_8\}$,
and are determined by finding minimal abductive explanations for $\mc{Q}(\nit{John})$. However, without considering the IC $I$.

Now, these explanations have to be examined at the light of the ICs. In this case, $E_1$ does satisfy $I$, but this is not the case for $E_2$. So, the
latter is rejected. As a consequence, the only admissible update is the deletion of $t_1$ from $D$.

The admissible (and minimal) view-updates could be used {\em to define} actual causes under ICs. In this case, and according to
Section \ref{sec:delp&cause}, the admissible view-update (under ICs) should be in correspondence, by definition, with an {\em admissible} and
minimal combination
of an actual cause and one of its
contingency sets. This would make $t_1$ the only actual cause (also counterfactual) for $\langle \nit{John}\rangle$ under $I$, which corresponds
with the result obtained following our direct definition. \boxtheorem
\end{example}

 \section{Conclusions}

In combination with the results reported in \cite{icdt15}, we can see that there are deeper and multiple connections between the areas
of QA-causality, abductive and consistency-based diagnosis, view updates, and database repairs. 
Abduction has also been explicitly applied to database repairs \cite{arieli}. The idea, again, is to ``abduce" possible repair updates that
bring the database to a consistent state. Connections between consistency-based and abductive diagnosis have been established, e.g. in \cite{ConsoleT91}.
Exploring and exploiting all the possible connections is matter of ongoing and future research.

\ignore{
We point out that database repairs are explicitly related to the view-update problem.
Actually, {\em answer set programs} (ASPs) \cite{asp} for database repairs \cite{2011Bertossi}  implicity repair the database by updating conjunctive combinations of intentional,
annotated predicates. Those logical combinations -views after all- capture violations of integrity constraints in the original database or along the (implicitly iterative) repair process
(a reason for the use of annotations).  }

\ignore{
Even more, in \cite{lechen}, in order to protect sensitive information, databases are explicitly and virtually ``repaired" through secrecy views that specify the
information that has to be kept secret. In order to protect
information, a user is allowed to interact only with the virtually repaired versions of the original database that result from making those views empty or
contain only null values. Repairs are specified and computed using ASP, and an explicit connection to prioritized attribute-based repairs \cite{2011Bertossi} is made \cite{lechen}.  }

\vspace{2mm}
\noindent {\small {\bf Acknowledgments:} \ Research funded by NSERC DG (250279), and
the NSERC Strategic Network on Business Intelligence (BIN).}

\vspace{-1mm}


\ignore{
\defproof{ Proposition \ref{pro:relp}}{ To show the membership to {\em  {NP}}: given a Datalog abduction $\mc{AP}$ and $h \in Hyp$, non-deterministically guess a subset $\Delta \subseteq\nit{Hyp}$, check if a) $h \in \Delta$ and b) $\Delta$ is an abductive diagnosis for $\mc{AP}$, then $h$ is relevant, Otherwise, it is not relevant. Obviously, (a) can be checked in polynomial. We only need to show that checking (b) is also polynomial. More  precisely, we need to show that $ \Pi \cup E \cup \Delta \models \nit{Obs}$ and $\Delta$ is subset-minimal. Checking whether $ \Pi \cup E \cup \Delta \models \nit{Obs}$ can be done in polynomial, because Datalog evaluation is polynomial time. It is easy to verify that to check the minimality of $\Delta$, it is good enough to show that for all elements $\delta \in \Delta$, $ \Pi \cup E \cup \Delta \not \models \nit{Obs}$. This is because positive Datalog is monotone. Therefore, relevance problem belongs to {\em  {NP}}.

We establish the hardness by showing that the combined complexity of the relevance problem for Propositional Horn Clause Abduction ( {PHCA}), shown to be {\em  {NP}}-complete in \cite{Friedrich90}, is an lower bound for the data complexity of the relevance problem for Datalog abduction.  {PHCA} is a tuple $P =(\nit{Var, \mc{H}, SD, \mc{O}})$, where $\nit{Var}$ is a finite set of propositional variables, $\mc{H} \subseteq \nit{Var}$ are
the individual Hypotheses, $\nit{SD}$ is a set of definite propositional Horn clauses, and $\mc{O} \subseteq \nit{Var}$  the observation, is a finite conjunction of propositions with $\mc{H} \cap \mc{O} = \emptyset$. An abductive diagnosis  for $P =(\nit{Var, \mc{H}, SD, \mc{O}})$ is a subset-minimal $\Delta \subseteq \mc{H}$ such that  $\Delta \cup \nit{SD} \models Obs$. It is known that deciding whether $h \in \mc{H}$ is relevant to $P$ (i.e., it is an element of an abductive diagnosis of $P$) is {\em {NP}}-complete.

We call a  {PHCA} 3-bounded if its rules are of the form  $\nit{true} \leftarrow$  or $a \leftarrow b_1, b_2, b_3$. It is clear that all  {PHCA}s can be converted to an equivalent 3-bounded  {PHCA}. Without loss of generality, assume $|\mc{O}|=1$.

For a 3-bounded  {PHCA}, we define a Datalog abduction  $\mathcal{AP}= \langle \Pi, E, \nit{Hyp}, \nit{Obs}\rangle$ where, $\Pi$ is
\begin{eqnarray*}
t(true)&\leftarrow&\\
t(x_0)&\leftarrow& t(x_1), t(x_2), t(x_3), r_i(x_0, x_1, x_2, x_3), \\
\end{eqnarray*}

\vspace{-.7cm}
and the input structure contains a ground atom $r_i(a, b_1, b_2, b_3)$ iff the rule $a \leftarrow b_1, b_2, b_3$ occurs in $\nit{SD}$. Furthermore, $\nit{Hyp}=\{ t(x)| \ x \in \mc{H} \}$ and $\nit{Obs}=\{t(x)| \ x \in \mc{O} \}$.

It is not difficult to verify that $h$ is a relevant hypothesis for a  {PHCA} $P$ iff $t(h)$ is relevant for the corresponding Datalog abduction $\mc{AP}$.   $\mc{AP}$ has a fixed program and the size of $\nit{SD}$ is pushed to the size of the input structure $\nit{EDB}$ i.e., combined complexity of the relevance problem for  {PHCA} is a lower bound for the data complexity of Datalog abduction. Therefore, relevance problem is {\em  {NP}}-hard.}
}

\ignore{
\defproof{Proposition \ref{pro:abdf&cfcaus}}{ Fist, assume $\tau$ is an actual for \nit{ans}. According to Definition \ref{def:querycause} (slightly modified for Datalog queries) there exists a contingency set  $ \Gamma \subseteq D^n$ s.t. $ \Pi \cup D \smallsetminus  \Gamma \models  \nit{ans}$ but $\Pi \cup D \smallsetminus \Gamma - \{t\} \not \models  \nit{ans}$. This implies that there exists a set $\Delta \subseteq D^n$ with $t \in \Delta $ s.t.   $ \Pi \cup \Delta \models \nit{ans}$. It is easy to see that  $ \Delta$ is an abductive diagnosis for $\mathcal{AP}^c$. Therefore, $t \in \nit{Rel}(\mathcal{AP}^c)$.

Second, assume $t \in \nit{Rel}(\mathcal{AP}^c)$. Then there exists a set $\mc{S}_k \in \nit{Sol}(\mathcal{AP}^c)=\{s_1 \ldots s_n\}$ such that $\mc{S}_k \models \nit{ans}$ with $t \in \mc{S}_k$. Obviously, $\nit{Sol}(\mathcal{AP}^c)$ is a collection of subsets of $D^n$. Pick a  set $\Gamma \subseteq D^n$ s.t., for all $\mc{S}_i \in \nit{Sol}(\mathcal{AP}^c)$ $i \not = k$, $\Gamma \cap \mc{S}_i \not = \emptyset$ and  $\Gamma \cap \mc{S}_k =\emptyset$. It is clear that $ \Pi \cup D \smallsetminus \Gamma -\{t\} \not \models \nit{ans}$ but  $\Pi \cup D \smallsetminus \Gamma \models \nit{ans}$. Therefore, $\tau$ is an actual cause for \nit{ans}.
To complete the proof we need to show that such $\Gamma$ always exists. This can be done by applying the digitalization technique to construct such $\Gamma$. Since all elements of $\nit{Sol}(\mathcal{AP}^c)$ are subset-minimal, then, for each $\mc{S}_i \in \nit{Sol}( \mathcal{AP}^c)$ with $i \not = k$, there exists a $t' \in \mc{S}_i$ such that $t' \not \in \mc{S}_k$.  Therefore, $\Gamma$ can be obtained from the union of difference between each $\mc{S}_i$ ($i \not = k$) and $\mc{S}_k$.
}
}

\ignore{
\defproof{Proposition \ref{pro:abdf&res}}{
Assume $N$ is a minimal cardinality set of necessary
hypothesis set of $\tau$. From the definition a necassry hypothesis set, it is clear that $\Gamma= N-\{t\}$ is a cardinality minimal contingency set for $\tau$ and the result is followed.
}
}

\ignore{
\defproof{Proposition \ref{pro:causeSubSDP}}{ The results is simply follows from the definition of an actual cause.  Assume a set $D' \subset D$ and $t \in  D \smallsetminus D' $.  If $D'$ is a solution to a source minimal side-effect deletion-problem then $\tau$ is an actual cause for $a$ with contingency  $D \smallsetminus D' -\{t\}$. Likewise, removing each actual cause for $a$ together with one of it S-minimal contingency set is a solution to source minimal
side-effect deletion-problem.
}
}

\ignore{
\defproof{Proposition \ref{pro:cp}}{  To show the membership to \nit{{NP}}:  non-deterministically guess a subset $\Gamma \subseteq D^n$, return yes if $D\cup  \Gamma \cup \{t\} \not \models \mc{Q}(\bar{a})$ and $\Gamma \cup \{t\} \models \mc{Q}(\bar{a}) $.  Otherwise no. Checking the mentioned conditions consists of  two Datalog query evaluations that is polynomial in data. The \nit{{NP}}-hardness is obtained by the reduction from the relevance problem for Datalog abduction to causality problem provided in \ref{pro:ac&rel}.}
}

\ignore{
\defproof{Proposition \ref{pro:causefromviewII}}{ The \nit{{NP}}-hardness is obtained by the reduction from the view-side-effect free problem for Datalog abduction to  {vc-}cause problem provided in \ref{pro:vc&view}. Membership to \nit{{NP}} can be shown similar to that of \ref{pro:cp} }
}

\end{document}